\documentclass[pra,twocolumn,amsfonts,amssymb,amsmath,floatfix,floats,a4paper]{revtex4-1}

\usepackage{graphicx}
\usepackage{xcolor}
\usepackage{amsmath}

\begin{document}

\title{Comment on ``Weak values and the past of a quantum particle''}
\date{May 2023}

\author{Lev Vaidman}

\affiliation{Raymond and Beverly Sackler School of Physics and Astronomy\\
 Tel-Aviv University, Tel-Aviv 6997801, Israel}
\begin{abstract}
In a recent paper, Hance, Rarity and Ladyman [Phys. Rev. Res. {\bf 5}, 023048 (2023)] criticized recent proposals connecting weak values and the past of a quantum particle. I argue that their conclusion follows from a conceptual error in understanding the approach to the past of the particle they discuss.
\end{abstract}

\maketitle

Hance, Rarity and Ladyman (HRL) \cite{HRL} discuss the connection between the presence of a quantum particle in the past and weak values, the topic I introduced in 2013 \cite{past}. They claim to analyze it according to my definition and also according to an alternative approach \cite{Cheshire}. In this Comment I argue that there is a conceptual error in the HRL paper in presentation of these approaches and consequently, their conclusion \begin{quote} ``that these approaches specifically are not useful for helping identify the past path of quantum particles'' \end{quote} misses the target.

The conceptual error of the HRL paper is that according to their presentation, the discussed approaches argue for the existence of an independent ontological concept of the presence of a pre- and postselected particle. Perhaps I should refrain from discussing the alternative approach \cite{Cheshire}, but I can say that this is definitely not true for my approach. The definition of the ``presence of the particle'' in \cite{past} is operational: {\it the particle was where it left a (weak) trace}.
Therefore, to ``identify the past path of quantum particles'' is to find the locations where they left a trace. The weak values of the local operators are a useful tool for calculating these local traces.
There is a controversy about the faithfulness of this method, but HRL mainly criticise the connection between weak values and hypothetical ``particle presence'' and not between weak values and weak traces, which can be calculated using standard quantum mechanics.
HRL write
\begin{quote}
``These approaches simply assume the particle was present wherever the weak value of an operator containing the spatial projection operator is nonzero.''
\end{quote}
The approach \cite{past} {\it defines} that the particle was present where it left a trace.
The purpose of this Comment is to clarify the approach by pointing out several misconceptions in the HRL paper.

HRL write (in the Introduction) about ``weak values only being defined over ensembles''. I, as co-author of the original paper which introduced weak values \cite{AAV}, disagree with this statement. It is correct that we usually need an ensemble to observe a weak value, but nothing prevents us from defining it for a single system \cite{beyond}.

 HRL also discuss disturbance in weak measurements. Apparently, they attach a weak value to a weak measurement. We do need to know the results of the preselection measurement and the postselection measurement, but the discussion of the presence of the particle between these measurements does not require weak measurement: the environment ``measures'' the weak value by being disturbed. The particle is also disturbed by the environment; the weak values are then modified and it is a subtle issue when this can or cannot be neglected \cite{PNAS}.

The definition of presence based on a weak trace requires the existence of all possible types of local interactions with the environment. These interactions must be non-vanishing but can be arbitrarily small. Their purpose is to serve as a reference to the trace left on the environment in the discussed experiments relative to a hypothetical experiment with a well-localized particle in the same location. Then I
do agree with
\begin{quote}
``the existence of at least one operator formed from
the product of the spatial projection operator for a location and
some other operator, with a nonzero weak value, is both a necessary
and a sufficient condition for particle presence at that
location''.
\end{quote}
It follows from the fact that nonvanishing local interactions ensure the first-order trace in the environment.

Note that in optical interferometric experiments we always have a finite interaction of the photon with mirrors.  (The HRL energy exchange estimate $10^{-33}$, has to be replaced by considering a much larger momentum exchange of every photon bouncing off a mirror. The amplitude of the orthogonal component of the quantum state of the mirror due to the bouncing photon in the same experiment is of the order $10^{-17}$.)

I agree with the HRL view that
\begin{quote}
``any attempt
to form a definition of presence for quantum particles should
correspond to our intuitions about classical presence, unless
we have a good reason for it to deviate from this.

The classical conception of a particle presence—being
present at a certain place at a certain time—can be characterized
as follows:

(i) Every particle is located in space at all times.

(ii) Particles cannot be on more than one path simultaneously.

(iii) Particle trajectories are continuous (or at least as continuous as space is) so particles cannot get from one place to
another without passing through the space in between.

(iv) Particles interact with other objects and/or fields local to their location.

(v) If a particle is on a path at a given time, and that path is within some region, then the particle is also located in that
region at that time.

(vi) If a particle’s property is at a location, the particle must be at that location too.''
\end{quote}

My attempt for a new definition came exactly when I found ``a good reason''. In the nested Mach-Zehnder interferometer \cite{V07} there is a contradiction between (iii) and (iv) .
The traces left on the environment that provide evidence of particle interactions have disconnected parts. We cannot retain all the classical characterizations of presence in the quantum world, so in \cite{past} I abandoned (iii) and adopted (iv) as the definition.

My definition allows for keeping all other properties, although there is a very subtle and paradoxical situation about property (ii). In the nested Mach-Zehnder interferometer the particle {\it is} present in every one of the three arms at the same time, but it is not present in any two (or three) arms simultaneously. To be present in a particular location is to leave a trace there, and the quantum states of the environment in all three arms are changed, so the particle was in three places. However, the traces in the environment are entangled such that orthogonal components of the local states of the environment, which provide evidence of the presence in every location, are entangled with undisturbed states in all other locations. Thus, there is no trace corresponding to {\it simultaneous} presence in different locations and one can claim that (ii) is fulfilled. A similar paradoxical situation arises in the Hardy paradox \cite{Hardy,HardyV,Product} which describes a pre- and postselected system of two particles: an electron was present in one arm and a positron was present in another, but the particles were not present in these arms simultaneously.

There is no similar difficulty with (v), since the definition is that if anywhere in the region there is a non-vanishing trace, the particle was in this region.  Note that due to the unavoidable momentum exchange of the photon with mirrors, there is no such thing as ``undisturbed inner interferometer'' discussed by HRL.
Contrary to the HRL claim, the fact that traces might have various properties (e.g. sign) is not neglected in the two-state vector formalism and asking specific questions like: Was the photon in two places together? leads to paradoxical answers, as in the discussion of (ii), see \cite{ChinaCom}.

I am puzzled by the HRL claim about the inconsistency of \cite{Peled}. Why ``would we expect a particle to necessarily have a non-zero weak value for the
spatial projection operator for any path along which it travels'' when interactions with other degrees of freedom lead to the trace?

I want to repeat that I disagree with Section VI of HRL, my weak value is defined for a single system. The fact that the weak value of the velocity of a particle can be larger than the speed of light (see Sec. VIII of \cite{AV90}) does not contradict the special theory of relativity. The experiments involve postselection and their low probability of success prevents a superluminal change in the probability of finding a quantum particle.

I also disagree with the claim of Sec. VII of the HRL paper, according to which the weak value approach is intended to show that some quantum protocols ``are not as `spooky' as they appear''. The weak value approach helps to find quantum protocols which are ``spooky'' if analyzed in classical terms. My papers based on the weak value approach cited by HRL \cite{57,58,59,60,62} do not try to remove paradoxical features. Instead, these papers try to correct erroneous claims about alleged counterfactual communication. In particular, HRL are correct in their weak values analysis of the protocol described in \cite{Laws}  and shown on their Fig.~2. The photons reaching detector $D_0$ were not present at Bob's site according to the weak trace criterion (all weak values of local operators on Bob's site vanish). However, there is no contradiction with the approach because, in this case, Bob's communication with Alice fails. The click at $D_0$ means that the photon did not perform any test of the presence or absence of Bob's shutter because the probability of this click does not depend on Bob's actions, see \cite{Counterport}.

Finally, let me comment on the concluding sentence of HRL
\begin{quote}
``we have shown that weak value approaches to the path of a particle do not contribute any new physics—the assumption of a connection between particle presence and weak
values does not give us anything testable.''
\end{quote}
First, weak values, as all other concepts and results of the two-state vector formalism are fully consistent with the standard formalism of quantum mechanics, so neither new physics, i.e., a deviation from the Schr\"odinger equation, nor introducing some new ontology, is proposed. My approach introduces new concepts (which I believe are useful), in particular, the local presence of a pre- and post-selected particle {\it defined} by the local trace it leaves on the environment. The formalism predicts that these traces can be found based on finite weak values of local operators, and this statement is definitely testable.

This work has been supported in part by the U.S.-Israel Binational Science Foundation (Grant No. 735/18) and by the Israel Science Foundation Grant No.~2064/19.

\bibliographystyle{unsrt}

\end{document}